# A Global View of Standards for Open Image Data Formats and Repositories

Jason R. Swedlow[1*], Pasi Kankaanpää[2], Ugis Sarkans[13], Wojtek Goscinski[3], Graham Galloway[5], Ryan P. Sullivan[6], Claire M. Brown[7], Chris Wood[8], Antje Keppler[9], Ben Loos[10], Sara Zullino[11], Dario Livio Longo[12], Silvio Aime[11], Shuichi Onami[14*]

[1]Divisions of Computational Biology and Gene Regulation and Expression, School of Life Sciences, University of Dundee, Dundee, UK

[2]Turku BioImaging, Åbo Akademi University and University of Turku, Turku, Finland; Euro-BioImaging ERIC, Turku, Finland

[3]Monash eResearch Centre, Monash University, Australia

[5]National Imaging Facility, The University of Queensland, Queensland, Australia

[6]Microscopy Australia, The University of Sydney, Sydney, Australia

[7]Advanced BioImaging Facility (ABIF), McGill University, Montreal, Canada; and Canada BioImaging

[8]Laboratorio Nacional de Microscopía Avanzada, Instituto de Biotecnología, Universidad Nacional Autónoma de México, Cuernavaca, México.

[9]Euro-BioImaging Bio-Hub, European Molecular Biology Laboratory, Heidelberg, 69117 Germany

[10]Department of Physiological Sciences, Stellenbosch University, Stellenbosch, South Africa

[11]Molecular Imaging Center, Department of Molecular Biotechnology and Health Sciences, University of Torino, Italy; Euro-BioImaging ERIC, Torino, Italy

[12]Institute of Biostructures and Bioimaging (IBB), National Research Council of Italy (CNR), Torino, Italy

[13] European Molecular Biology Laboratory, European Bioinformatics Institute (EMBL-EBI), Wellcome Genome Campus, Cambridge, UK

[14]RIKEN Center for Biosystems Dynamics Research, Kobe, 650-0047, Japan

Corresponding authors: jrswedlow@dundee.ac.uk, sonami@riken.jp

**Abstract**

Biological and biomedical imaging datasets record the constitution, architecture and dynamics of living organisms across several orders of magnitude of space and time. Imaging technologies are now used throughout the life and biomedical sciences to achieve discovery and understanding of biological mechanisms in the basic sciences as well as assessment, diagnosis and therapeutic intervention in clinical trials and animal and human medicine. The universal application and use of imaging raises an important question and opportunity: what is the value and ultimate destination of biological and medical imaging data? In the last few years, several informatics and data science technologies have matured sufficiently so that routine publication of these datasets is now possible. Participants in Global BioImaging from 15 countries and all populated continents have agreed on the need for recommendations and guidelines for the establishment of image data repositories and the formats they use for delivering data to the global scientific community. This white paper presents a shared, global view of criteria for these common, globally applicable guidelines and provisional proposals for open tools and resources that are available now and can provide a foundation for future development.

**Introduction**

Imaging is now used globally as a method of quantitative measurements of biological and biomedical structure, constitution and dynamics in the life and biomedical sciences. Imaging technology is rapidly evolving with new modalities and applications appearing that enable new insights and discoveries [1,2]. This rapid innovation presents challenges at several different but interdependent levels. Sourcing and retaining expert research technology professionals ("imaging scientists"), providing initial and ongoing training in advanced technologies, rapid dissemination and easy access to new innovative methods and applications, and data management and analysis are global issues that are experienced by research labs and institutions across all sectors that use imaging, including academic and industrial research. Global BioImaging (https://globalbioimaging.org) was founded to meet these challenges, and wherever possible use the spirit of cooperation and collaboration across international boundaries to share and define best practice, develop common imaging and data standards, and develop shared world-class training programmes and tools for imaging scientists.

Global Bioimaging's 'Exchange of Experience' meetings have repeatedly discussed the need for standards for image data formats and public data resources. This issue has become critical with the emergence of several new public data resources. Moreover, with the rapid innovations in light-sheet microscopy, multiplex tissue imaging, spatial profiling of single cell transcriptomes, mass spectrometry-based imaging, correlative imaging techniques, molecular imaging and several others, data complexity and dimensionality are increasing, which makes the need for open, common methods for recording imaging metadata even greater. It is essential we address this challenge so that individual labs, large multi-center projects and public data resources have the solutions they need to enable analysis, sharing and publication of this new generation of datasets. Global Bioimaging' partners (see Table 1 for the partners participating in this effort) have observed these challenges emerging across all boundaries of geography and scientific domain, and therefore have come together to try to identify universally relevant solutions.

| Entity | Location | URL | Description |
|---|---|---|---|
| Euro-BioImaging ERIC | Europe (1) | eurobioimaging.eu | European Research Infrastructure Consortium |
| Advanced Bioimaging Support | Japan | www.nibb.ac.jp/abis | National Imaging Infrastructure Consortium |
| Canada BioImaging | Canada | www.canadabioimaging.org | National Imaging Consortium |
| Bioimaging North America | USA, Mexico, Canada | www.bioimagingna.org | TransNational Imaging Consortium |
| SINGASCOPE | Singapore | | National Imaging Infrastructure Consortium |
| Microscopy Australia | Australia | micro.org.au | National Imaging Infrastructure Consortium |
| National Imaging Facility | Australia | www.anif.org.au | National Imaging Infrastructure Consortium |
| National Laboratory for Advanced Microscopy | Mexico | lnma.unam.mx/wp/ | National Imaging Consortium |
| South Africa BioImaging | South Africa | | National Imaging Consortium |
| India BioImaging Consortium | India | bioimaging.bio.iisc.ac.in | National Imaging Consortium |
| Open Microscopy Environment (2) | United Kingdom | openmicroscopy.org | Open data and software consortium |
| Systems Science of Biological Dynamics (2) | Japan | ssbd.qbic.riken.jp | National open data and software project |
| Australian Characterisation Commons at Scale and Characterization Virtual Laboratory (2) | Australia | cvl.org.au | National cross-modality imaging informatics infrastructure project |

Table 1. Global BioImaging partners and participating national and international initiatives (2) (status: October 2020). (1): Euro-BioImaging ERIC is a distributed research infrastructure with its statutory seat located in Finland. EMBL hosts the community-specific section for biological imaging as well as general data services. Italy hosts the community-specific section for medical imaging.

**Target Audiences for Global Bioimaging Recommendations**

For construction and dissemination of recommendations by the Global Bioimaging community, the target audiences for our recommendations are a broad range of constituencies and community members. We aim to support and ideally influence imaging scientists - central facility staff and managers who deliver technical know-how and best practice to their experimental science colleagues. However, Global BioImaging also seeks to influence journal editors and funders, who have an important role in defining policy, practice and implementation. Journals have repeatedly contributed to the use of open data standards by requiring that papers submitted for publication adopt domain-specific data deposition standards (deposition of sequence data in ArrayExpress or European Nucleotide Archive, structural data in the Protein Data Bank, etc.). Funders contribute by conditioning funding awards on the use and adoption of data standards, and where appropriate, the deposition of datasets in open repositories. Global BioImaging's recommendations are constructed so that they can be easily appreciated and incorporated by a wide cross-section of the scientific community.

**Serving a Diverse Collection of National Communities**

Global Bioimaging uses its international training programs and 'Exchange of Experience' meetings to share expertise and know-how between different imaging communities and domains while also developing an understanding and appreciation for the different levels of funding, installed technology, and scientific priorities in different countries and international regions. A key theme in the 'Exchange of Experience' meetings is constructing recommendations for defining and adopting standards (for image data, quality management, impact management, training curricula for facility staff, etc.), which may be of critical and multifaceted relevance, especially in respect of new or developing bioimaging communities. We believe that, particularly in resource-limited environments, international guidelines and standards will encourage the design of high quality experiments and ensure results are published with global visibility and impact, but will also drive the implementation of substantial bioimaging components that shape and contribute to MSc and PhD studies. In addition, there is a significant responsibility held by universities to ensure that the teaching and training around technologies and tools that are up to date and accurate. This is especially important in countries or regions with a skills shortage. Connecting developing groups to a global community that collectively builds and shares best practices and training programmes will accelerate the growth of developing science communities and minimize risk of poor practices in data generation, handling and analysis. We also aim to support the strategic focus and investment by funders by providing information on global trends and requirements for data handling, processing and analysis tools.

Moreover, Global BioImaging recommendations will support the growing international network of research infrastructure providers, who are increasingly responsible for guidance on best practice, developing and implementing processes, or making decisions on behalf of, and in partnership with, their user community. This includes both facilities based on physical instrumentation for capturing scientific experiments, and eResearch or cyberinfrastructure facilities who are responsible for the data management and analysis environments. Ultimately our goal is to leverage the richness and complexity of image data for new directions in research and training, for example for the application of new artificial intelligence methods for object recognition or correlation of multi-scale data sets.

For all these reasons, Global BioImaging partners are constructing international recommendations alongside their participating bioimaging communities, representing scientific communities from around the globe, and including imaging scientists and staff, universities and research institutions, national funders and science-policy makers.

**Building on Previous Experience**

The huge range of imaging modalities and applications reflects the incredible spread and dominance of imaging as a critical scientific technology in the physical, biological and biomedical sciences. This diversity demonstrates the power of imaging but also creates several challenges. In particular, the huge number of data formats that are used across many different modalities inhibits access to and exchange of datasets among scientists in collaborative projects, between different imaging applications and across research domains.

It is impractical to suppose or recommend that a single data format can satisfy the wide range of imaging applications used by the global community of imaging scientists. Thus, we have developed a series of specifications and recommendations for potential standards that Global BioImaging members, imaging scientists, journals, funders and institutions may adopt and use in the future. These recommendations are built upon the successful use of standards in various imaging communities, e.g., DICOM (Digital Imaging and Communications in Medicine, [http://medical.nema.org](http://medical.nema.org) ), OME-TIFF [3], imzML [4], Nifti ([https://nifti.nimh.nih.gov](https://nifti.nimh.nih.gov)), and many others. These various community standards have had different levels of success depending on the quality of implementation and ongoing maintenance of the format. They present an opportunity to learn from past successes and failures and provide the basis for a strong set of recommendations for defining and adopting standards for the worldwide bioimaging community, and in doing so, may provide guidelines that respect the different levels of development in different communities and countries. In the sections below we detail our current level of experience and recommendations for implementing and adopting standards for imaging data.

**Recommendations for Data Format Standards**

In the following we outline the characteristics of useful, usable data standards. These guidelines can be used by scientists, infrastructure providers and their personnel, funders and journal editors to assess the utility of data standards proposed by scientific groups, national programmes or transnational collaborations. These recommendations reflect the requirements that are increasingly being adopted by other communities [5].

1.        *Openness*

Any proposed data format must be openly available, supported by openly accessible, versioned, and editable specification(s), implementations, and documentation. Specifications and other related documents must be easily accessible from a URL or other publicly available on-line resource, following the FAIR specification—Findable, Accessible, Interoperable and Reusable—formulated by the Force11 group (https://www.force11.org/group/fairgroup/fairprinciples ). It is insufficient for documents and specifications to only be supplied on demand.

2.        *Implementation*

Any proposed format should be supported by open source, publicly available software, with well defined specifications, that provide read and write functions for the format, preferably in multiple, community-adopted programming environments (e.g., Java, Python, C++, etc). These implementations should include an application programming interface (API), and an open source reference implementation, so they can be easily adopted and included in 3$^{rd}$ party software. It is useful for the read functions to be incorporated into a validator, an application that can be used to read a file and assess how well it adheres to the standard. Software libraries that meet these requirements will serve as *reference implementations* for these formats, i.e., public tools that implement community-agreed guidelines and specifications and can be adopted and used by the broad target audience defined by Global BioImaging.

3.        *Examples*

Usage and adoption of a proposed data format standard will be catalysed by openly available examples—real data stored in the format. These are useful references for anyone wishing to adopt and use the format, and also can serve as tools for testing and validating software that reads and/or writes the format. For each version of the format specification, up-to-date examples should be provided.

4.        *Licensing*

All data standard resources including documentation, specifications, implementations, and example data sets should be licensed using an appropriate community-agreed license (one example are the Creative Commons licenses, e.g., CC0 or CC-BY). Licenses that forbid commercial use often inhibit adoption by industrial research labs and commercial technology providers and should be avoided. Software for reading/writing data formats should be licensed under a permissive software license, e.g., BSD, MIT, or similar in order to promote adoption by users from across the bioimaging community.

*5.        Data Types*

There are many different data types covering a multitude of different applications, domains, imaging modalities and spatial and temporal scales. Any proposed standard will likely only cover one or at most a few applications or domains. The expected types of data, supported by the standard, should be stated clearly in any documentation. In addition, the types of data supported, for example metadata related to experimental or case manipulations, image data acquisition, data processing, and analytic outputs should be clear, easy to understand for any user, and usable for search and data management applications.

*6.        Governance or change management*

For a scientific standard to stay relevant whilst ensuring transparency, it needs a mechanism or structure for decision-making and change management. Due to the varying types of standards, their reach, and differences across their adoptive community, a governance or change management policy and process could take many forms. The most critical attributes are transparency and strong community engagement.

*7.        Adoption*

For a standard to be considered suitable it should be adopted beyond an individual research laboratory, institution, or geographic locale.

**Data Repositories**

Commonly shared open datasets have repeatedly proven to be essential for the development of analytic and processing tools for data across the sciences. Open science initiatives are becoming more widely accepted by the scientific community and open access to research data is now often required by private, national and transnational funding agencies [6]. In the life and biomedical sciences, the commitment of the genomics community to rapidly publish genomic sequence data [7] was the basis of the development and growth of the modern field of bioinformatics. Global BioImaging aims to catalyze a similar development of bioimage informatics and data analytics by encouraging and supporting the construction, sustainability and continuous availability of repositories for imaging data.

Imaging datasets are rich, heterogeneous and often quite large. Until recently, most image data repositories published datasets from single projects, making large strategic datasets available for query and download. However, in the last 10 years, several repositories have appeared that integrate datasets from independent peer-reviewed studies enabling datasets from electron microscopy, high content screening, multi-dimensional fluorescence microscopy, histology, magnetic resonance imaging, positron emission tomography, ultrasound and several different modalities for whole tissue or organism imaging to be published and accessed online, usually through a web browser-based interface, and sometimes through appropriate APIs (see Table 2).

**Recommendations for Open Access Image Data Repositories**

Table 2 lists several image data archives and AVDBs used by Global BioImaging's scientists. The appearance and growth of these and other resources demonstrates that many of the barriers for managing and publishing large collections of images have been solved. They provide components of a coherent, connected ecosystem for publishing and integrating bioimaging data. We have therefore defined key, specific recommendations that should be implemented to ensure this momentum continues and preferably grows.

| Data Type | Utility & Impact | Types of Users/Applications | Examples of Public Resources | References |
|---|---|---|---|---|
| **Correlative light and electron microscopy** | Link functional information across spatial and temporal scales with ultrastructural detail | Cell biologists, structural biologists and modellers: structural models that span spatial and temporal scales | EMPIAR (https://www.ebi.ac.uk/pdbe/emdb/empiar); | 8 |
| **Cell and tissue atlases** | Construction, composition and orientation of biological systems in normal and pathological states. | Educational resources; Reference for construction of tissues, organisms, health scientists | Allen Brain Atlas (https://www.brain-map.org); Allen Cell Explorer (https://www.allencell.org/); Human Protein Atlas (https://www.proteinatlas.org); Mitotic Cell Atlas (https://www.mitocheck.org/mitotic_cell_atlas); The Whole Brain Atlas (http://www.med.harvard.edu/AANLIB/home.html); eMouse Atlas Project (https://www.emouseatlas.org/) | 9, 10, 11, 12, |

| Benchmark and Reference datasets | Standardised test datasets for new algorithm development | Algorithm developers; Testing systems | EMDataBank (http://www.emdatabank.org); BBBC (https://data.broadinstitute.org/bbbc); IDR (https://idr.openmicroscopy.org); CELL Image Library (http://www.cellimagelibrary.org); Single Molecule Localization Microscopy (http://bigwww.epfl.ch/smlm/datasets) | 13, 14, 15, 16, |
|---|---|---|---|---|
| Systematic Phenotyping | Comprehensive studies of cell structure, systems and response | Cell biologists, physiologists, Queries for genes or inhibitor effects | MitoCheck (http://www.mitocheck.org); SSBD (http://ssbd.qbic.riken.jp); IMPC (www.mousephenotype.org); PhenoImageShare (http://www.phenoimageshare.org/) | 11, 17, 18, 19 |
| Whole organ and Systems | Studies of whole tissues, animals or humans | Radiologist, physicians, algorithm developers, cell and systems biologists | Human Connectome Project (http://www.humanconnectomeproject.org/) The Cancer Imaging Archive (TCIA) (https://www.cancerimagingarchive.net/) Alzheimer's Disease Neuroimaging Initiative (ADNI) (http://adni.loni.usc.edu/ ) European Population Imaging Infrastructure (EPI2) (http://populationimaging.eu/) The BRAINS Imagebank (http://www.brainsimagebank.ac.uk/) Japanese Society of Radiological Technology (JSRT) (http://db.jsrt.or.jp/eng.php) | 20, 21, 22, 23, 24 |

**Table 2. Examples of Public Image Data Archives and AVDBs.** This table is exemplary and is not a comprehensive survey of all public imaging data resources.

1.  *Metadata Specifications for Submission*

The value of published imaging datasets can only be realised if they are accompanied by metadata that describe type and state of sample, experimental manipulations, imaging technology, conditions and probes, and any analytic results derived from the data. The value of capturing these metadata as completely as possible has to be weighed against the practicality of capturing experimental and analytic outputs from research laboratories. As noted above, there are several established metadata-rich formats (e.g., DICOM, OME-TIFF, NifTi and others), but the complexity of case, tissue, disease, sample and imaging modality metadata have defied full standardization, especially in the research setting-- innovative experiments and technologies often challenge previously used definitions and concepts [25]. New web-based metadata technologies like JSON-LD, which is now a formal specification from W3C, may provide a way to implement a flexible metadata spec in a common language. Nonetheless, in our experience, the easiest, most commonly used data format for research metadata is the spreadsheet, so public data resources will need to take a flexible, practical approach to capturing the broad range of metadata required to document and reproduce innovative experiments. Moreover, the increasing number of image data repositories may result in an equivalent number of metadata submission templates, causing confusion for data submitters and future data users.

The developing image data resources should engage with the bioimaging community to define, as much as possible, a common metadata specification that is shared across repositories, updated on a regular, predictable basis and relatively easy for data submitters to use, fill out and submit. The bioimaging community should collaborate to define consistent ontologies for metadata. As far as possible metadata should be harvested from the instrument, and at the time of acquisition. This will minimise any additional workload on the part of the researcher.

2.      *Components of the BioImage Data Ecosystem*
The collection, annotation, storage, integration and publication of biological datasets is well-established with many resources having reached maturity and stability. These existing resources serve as models that the imaging community can use to learn useful and successful design and construction patterns [26].

An approach that has proven successful in several other fields is to construct two separate data resources. The first, an *archive* or *repository*, holds and serves all data associated with publications, and stores data files and a limited amount of metadata. Data can be browsed, found using search indices and downloaded, but higher level annotation, integration and processing is not attempted, so that the archive can keep pace with the rate of data submissions. Archives hold datasets that are as close to the primary produced by the instrument as possible, and should be immutable. A second type of resource, an *added value database* (AVDB), incorporates datasets from the archive, performs curation and integration and seeks to enrich data and enable discovery with the datasets it holds ("reference datasets", see Ref [15]). The separation between the construction and operation of archives and AVDBs is critical to facilitate an efficient data intake workflow and also to allow curation at a sufficient level to enable data re-use and discovery.

Significant steps towards the establishment of a mature, usable bioimage data ecosystem have recently been achieved. Image databases that collect and curate multi-dimensional bioimaging data in electron microscopy, cell and tissue light microscopy, and several organ-specific resources as well as biomedical image data repositories are now funded, available and accepting and publishing terabyte-scale datasets (see Table 2). The launch of the BioImage

Archive (https://www.ebi.ac.uk/bioimage-archive/) by EMBL-EBI in July 2019 provided a central resource for the biological community and a common cross-domain foundation for existing and future added-value image databases such as those listed in Table 2.  Data pipelines from the BioImage Archive are being developed to connect to and feed AVDBs such as EMPIAR, and the Cell and Tissue IDR. Medical imaging communities are actively exploiting a dedicated image archive platform (XNAT, www.xnat.org) and developing tools for easy integrations with the BioImage Archive and other databases.  In the future, capabilities will be available to connect other AVDBs that aim to enhance the scientific value of the archived images through curation, integrative analysis and the development of new analytical methods for cross-interrogation and information retrieval among multi-domain AVDBs. Global Bioimaging strongly endorses these steps and looks forward to contributing to and deriving value from these public resources.

3.      *Requirements for AVDBs for Artificial Intelligence Applications*

As AVDBs grow and mature, the well-annotated datasets they hold may be valuable training datasets for advanced artificial intelligence (AI) applications, including tools that use deep learning. However, in discussions with members of GBI who run AVDBs, there is a shared sense that there aren't clear, definitive requirements for how training datasets should be constructed, how annotations ("labels") should be formatted, or which datasets should be prioritised for formatting for AI uses. We recommend that custodians of AVDBs work with AI experts to define these and other requirements in order to rapidly expand the usage of bioimaging datasets for AI applications. This should include standards for linking the imaging data to other relevant data from the same subject/sample, such as genetic data and biochemical/clinical/behavioral results.

Moreover, there are clearly strong opportunities for applying AI techniques to microscopy and imaging problems. Without community consensus across these attributes, we will impede the translation of AI image analysis techniques from laboratory to application, taking into account the growing demands for greater transparency of AI operative heuristics and legislation for the right to an explanation of algorithmic decision making.

4.      *Authentication for Submissions and Data Access*

As archives and AVDBs grow, the number of submissions they receive will increase, and the number of authors submitting datasets will also increase. This will inevitably raise an issue whereby authentication of author identity, affiliation and other critical information becomes an essential part of the data submission workflow. Several public, diverse identifier and authentication projects, including ORCiD (https://orcid.org/), Elixir Authentication and Authorization Infrastructure (https://www.elixir-europe.org/services/compute/aai), Life Science Authentication and Authorization Infrastructure (LS AAI) (https://tnc18.geant.org/getfile/4229), and Australian Access Federation (AAF https://aaf.edu.au/ ) are building identification policies and resolution systems to ensure all members of the scientific community are associated with a unique identifier and to provide services to resources like the imaging archives and AVDBs for user identification and authorization.

LS AAI is an extensive collaborative project where several life science research infrastructures have together defined requirements for a common AAI, developed under the overarching blueprint of the AARC (https://aarc-project.eu/). LS AAI is currently being implemented within the EOSC-Life project (http://www.eosc-life.eu/), and is foreseen to be widely used by the life science community in the future. In another example, the AAF provides a federated web-login service that allows researchers to access a broad variety of Australian research-focused web services through their University credentials. It is used for authentication to access gateways (CVL, Genomics Virtual Laboratory), repositories (Store.* and ImageTrove) and other resources.

We recommend that those involved in data services develop a task force to research current and ongoing work and standardised authentication practice and initiate proof of concept projects to assess the usage and usability of the various authentication systems that are coming on-line. In the long-term, a truly global identification and authentication could be  extended to identify instruments and the datasets they collect.

5. *Trustworthy Research Data Resources*

The complexity of acquisition techniques, experiments and the resulting research data is increasing, and this challenges data archives and AVDBs, and ultimately the ability to recreate experiments or reuse data. In response, there are emerging efforts to assess and declare the quality level of public data resources, using criteria of adoption of community standards, openness and sustainability.  These efforts are international and extend across a broad range of scientific domains. Examples include FAIRsharing.org, which provides a catalogue and characteristics of databases, data standards and other public resources [27] CoreTrustSeal's Core Trustworthy Data Repositories Requirements ( https://www.coretrustseal.org/), which provides a list of requirements that are deemed mandatory for a trustworthy data repository. In Australia the National Imaging Facility is building a trusted data resource to serve the needs of its national community [28]. In the European Union, EOSC-LIfe is constructing a trusted, sustainable open data resource infrastructure for the Life Sciences (https://www.eosc-life.eu/). These efforts aim to increase reproducibility and repeatability of experiments; increase researchers understanding the data; and make processing pipelines humanly transparent and increase data provenance.

6. *Human Identifiable Data*

A key issue for these resources are the methods and policies around identifiable data, and/or datasets derived from individuals or their biological material [29]. There are several active efforts to define guidelines for both ethical and best practice in the sharing and publication of these data.  For example, guidelines published by the Global Alliance for Genomics and Health (e.g., https://www.ga4gh.org/genomic-data-toolkit/regulatory-ethics-toolkit/) provide a useful, established framework for the developing bioimage data ecosystem.  As bioimage data resources will undoubtedly link to and/or integrate genomics and other datasets, their adoption of these existing guidelines will likely be the most sensible and efficient way to handle these precious datasets.

**Future Directions**

Looking forward, we see several challenges and opportunities for the current crop of imaging data formats and resources.  Most formats will not perform well in cloud-based storage technologies ("object storage") so new binary and metadata storage technologies will be required.  Whole tissue or body profiling projects, e.g. the Human Biomolecular Atlas Project [30], are creating datasets that far exceed the capabilities of the current generation of file formats and resources. Support for new types of metadata that integrate experimental protocols, organism metadata, common coordinate frameworks, analytic results and derived models are urgently required. Finally the application of ML-based models for object recognition and segmentation will require wholly new capabilities in data resources, so that well-annotated models can be published and shared.  We recommend that academic and commercial technology developers, funding agencies, and experimental and computational users of these resources specify and begin to construct the data technologies required for the next generation of imaging experiments.

**Conclusion**

Standardised data formats and public data resources are a critical "next step" for the fields of biological and biomedical imaging.  The appearance of several open data formats and data repositories has demonstrated that the technology and know-how exist to build these resources. The members of GBI agree that the next step is to drive adoption by all members of the scientific community, but in particular funders and journals who can mandate the use of open formats and data deposition as a condition of funding or acceptance of scientific publications. We have outlined the characteristics of standards that can be used by these critical stakeholders to assess the quality of proposed open formats and data repositories.

**Acknowledgements**

The authors thank Global BioImaging for providing an ample and useful environment at its Exchange of Experience meetings for the discussions that led to this paper.  Global BioImaging currently receives funding  from the Chan Zuckerberg Initiative. Between 2015 and 2018, Global

BioImaging project was funded by the Horizon 2020 Framework Programme of the European Union (project 653493).

# References

1. Swedlow, J. R. Innovation in biological microscopy: current status and future directions. *Bioessays* **34**, 333–340 (2012).
2. Jaffray, D. A., Das, S., Jacobs, P. M., Jeraj, R. & Lambin, P. How Advances in Imaging Will Affect Precision Radiation Oncology. *Int. J. Radiat. Oncol. Biol. Phys.* **101**, 292–298 (2018).
3. Linkert, M. *et al.* Metadata matters: access to image data in the real world. *J. Cell Biol.* **189**, 777–782 (2010).
4. Schramm, T. *et al.* imzML--a common data format for the flexible exchange and processing of mass spectrometry imaging data. *J. Proteomics* **75**, 5106–5110 (2012).
5. Martone, M. *et al.* International Neuroinformatics Coordinating Facility Review Criteria for Endorsement of Standards and Best Practices v. 1.0. *F1000Res.* **8**, (2019).
6. *Directive (EU) 2019/1024 of the European Parliament and of the Council of 20 June 2019 on open data and the re-use of public sector information*. https://eur-lex.europa.eu/legal-content/EN/TXT/?uri=CELEX:32019L1024.
7. Maxson Jones, K., Ankeny, R. A. & Cook-Deegan, R. The Bermuda Triangle: The Pragmatics, Policies, and Principles for Data Sharing in the History of the Human Genome Project. *J. Hist. Biol.* **51**, 693–805 (2018).
8. Iudin, A., Korir, P. K., Salavert-Torres, J., Kleywegt, G. J. & Patwardhan, A. EMPIAR: a public archive for raw electron microscopy image data. *Nat. Methods* **13**, 387–388 (2016).
9. Sunkin, S. M. *et al.* Allen Brain Atlas: an integrated spatio-temporal portal for exploring the central nervous system. *Nucleic Acids Res.* **41**, D996–D1008 (2013).
10. Uhlen, M. *et al.* Towards a knowledge-based Human Protein Atlas. *Nat. Biotechnol.* **28**, 1248–1250 (2010).
11. Cai, Y. *et al.* Experimental and computational framework for a dynamic protein atlas of human cell division. *Nature* **561**, 411–415 (2018).
12. Richardson, L. *et al.* EMAGE mouse embryo spatial gene expression database: 2014 update. *Nucleic Acids Res.* **42**, D835–44 (2014).
13. Lawson, C. L. *et al.* EMDataBank unified data resource for 3DEM. *Nucleic Acids Res.* **44**, D396–403 (2016).
14. Ljosa, V., Sokolnicki, K. L. & Carpenter, A. E. Annotated high-throughput microscopy image sets for validation. *Nat. Methods* **9**, 637 (2012).
15. Williams, E. *et al.* The Image Data Resource: A Bioimage Data Integration and Publication Platform. *Nat. Methods* **14**, 775–781 (2017).
16. Orloff, D. N., Iwasa, J. H., Martone, M. E., Ellisman, M. H. & Kane, C. M. The cell: an image library-CCDB: a curated repository of microscopy data. *Nucleic Acids Res.* **41**, D1241–D1250 (2013).
17. Tohsato, Y., Ho, K. H. L., Kyoda, K. & Onami, S. SSBD: a database of quantitative data of spatiotemporal dynamics of biological phenomena. *Bioinformatics* **32**, 3471–3479 (2016).
18. Cacheiro, P., Haendel, M. A., Smedley, D. & International Mouse Phenotyping Consortium and the Monarch Initiative. New models for human disease from the International Mouse Phenotyping Consortium. *Mamm. Genome* **30**, 143–150 (2019).
19. Adebayo, S. *et al.* PhenoImageShare: an image annotation and query infrastructure. *J. Biomed. Semantics* **7**, 35 (2016).
20. Van Essen, D. C. *et al.* The Human Connectome Project: a data acquisition perspective. *Neuroimage* **62**, 2222–2231 (2012).
21. Clark, K. *et al.* The Cancer Imaging Archive (TCIA): maintaining and operating a public information repository. *J. Digit. Imaging* **26**, 1045–1057 (2013).

22. Weiner, M. W. *et al.* Impact of the Alzheimer's Disease Neuroimaging Initiative, 2004 to 2014. *Alzheimers. Dement.* **11**, 865–884 (2015).

23. van den Bouwhuijsen, Q. J. A. *et al.* Determinants of magnetic resonance imaging detected carotid plaque components: the Rotterdam Study. *Eur. Heart J.* **33**, 221–229 (2012).

24. Job, D. E. *et al.* A brain imaging repository of normal structural MRI across the life course: Brain Images of Normal Subjects (BRAINS). *Neuroimage* **144**, 299–304 (2017).

25. Huisman, M. *et al.* Minimum Information guidelines for fluorescence microscopy: increasing the value, quality, and fidelity of image data. (2019).

26. Ellenberg, J. *et al.* A call for public archives for biological image data. *Nat. Methods* **15**, 849–854 (2018).

27. Sansone, S.-A. *et al.* FAIRsharing as a community approach to standards, repositories and policies. *Nat. Biotechnol.* **37**, 358–367 (2019).

28. Mehnert, A. J. *et al.* Putting the Trust into Trusted Data Repositories: A Federated Solution for the Australian National Imaging Facility. *IJDC* **14**, 102–113 (2019).

29. Knoppers, B. M. & Greely, H. T. Biotechnologies nibbling at the legal 'human'. *Science* **366**, 1455–1457 (2019).

30. HuBMAP Consortium. The human body at cellular resolution: the NIH Human Biomolecular Atlas Program. *Nature* **574**, 187–192 (2019).